\documentstyle[12pt]{article}

  \let\p=\pi

\def\0{\over } \def\1{\vec }     \def\2{{1\over2}} \def\4{{1\over4}}
\def\5{\bar }  \def\6{\partial } \def\7#1{{#1}\llap{/}}
\def\8#1{{\textstyle{#1}}}       \def\9#1{{\bf {#1}}}
 \def\llp{\hbox to 0pt{\hss /\hskip1.5pt}}
\def\llo{\hbox to 0.2pt{\hss /}} \def\llq{\hbox to 0pt{\hss /\hskip0.5pt}}
\def\so{\supset\hbox to 0pt{\hss $\displaystyle -$\hskip1pt}}

\def\<{\langle } \def\>{\rangle }

\let\nn=\nonumber
\def\bea{\begin{eqnarray}} \def\eea{\end{eqnarray}}
\def\beann{\begin{eqnarray*}} \def\eeann{\end{eqnarray*}}
\def\beq{\begin{equation}} \def\eeq{\end{equation}}

\newcommand{\Dl}{D\!\!\!\!\!\raisebox{1.5ex}{$\leftarrow$}}
\newcommand{\Dls}{D\!\!\!\!\!\raisebox{1.5ex}{$\leftarrow$}\!\!\!\!\!/}
\newcommand{\Ds}{D\!\!\!\!/}
\newcommand{\ls}{l\!\!/}

\newcommand{\pdl}{\partial\!\!\!\!\raisebox{1.5ex}{$\leftarrow$}}
\newcommand{\pdls}{\partial\!\!\!\!\raisebox{1.5ex}{$\leftarrow$}\!\!\!\!\!/}
\newcommand{\pds}{\partial\!\!\!/}
\renewcommand{\p}{{\scriptscriptstyle \|}}

\date{}
\title{
{\large\rm DESY 95-208}\hfill{\large\tt ISSN 0418-9833}\\
{\large\rm SLAC-PUB-7064}\hfill\vspace*{0cm}\\
{\large\rm December 1995}\hfill\vspace*{2.5cm}\\
Semiclassical Approach\\ to Structure Functions at Small x}
\author{W. Buchm\"{u}ller\\
{\normalsize\it Deutsches Elektronen-Synchrotron DESY, 22603 Hamburg, Germany}
\\[.2cm]
and\\[.2cm] A. Hebecker\\
{\normalsize\it Stanford Linear Accelerator Center, P.O.~Box 4349, MS 81,
CA 94309, USA}
\vspace*{2cm}\\
}
\begin{document}

\setlength{\baselineskip}{18pt}
\maketitle
\begin{abstract}
\noindent
Inclusive and diffractive structure functions for electron-proton scattering
are calculated in a semiclassical approach at large momentum transfer
$Q^2$ and small values of the scaling variable $x$.
The basic process is the production of a quark-antiquark pair in the
colour field of the proton. The structure functions are expressed
in terms of Wilson lines along the classical
trajectories of quark and antiquark passing through the colour field,
and their covariant derivatives. Based on some rather general assumptions on
properties of the colour field, inclusive and diffractive structure
functions are evaluated in terms of four field dependent constants.
\end{abstract}
\thispagestyle{empty}
\newpage

\section{Introduction}
The events with a large gap in rapidity, observed in deep inelastic
electron-proton scattering at HERA \cite{zeus,h1}, represent a puzzling
phenomenon. The separation of a colour neutral cluster of partons
from the proton, which then fragments independently of the proton remnant,
is a non-perturbative process. As the data show, the cross section for
these ``diffractive'' events is not suppressed at large values of $Q^2$
relative to the inclusive cross section. All this is difficult to understand
in the framework of perturbative QCD and the parton model, which
has been so successful
in describing many phenomena of strong interactions at short distances.

Various phenomenological models have been proposed which can account
for some aspects of the rapidity gap events. They combine ideas from
the quark-parton model and vector meson dominance \cite{bj},
Regge theory and perturbative QCD \cite{ingel}-\cite{frank}, or use
effective lagrangians together with vector meson dominance \cite{buch}.
In a recent paper \cite{heb} we have shown
that the global properties of the
rapidity gap events can be understood based on electron-gluon scattering
as underlying partonic process, provided a soft ``colour rotation''
\cite{nacht1}
transforms the produced quark-antiquark pair into a colour singlet.
Based on the same idea a detailed description of the final state has been
developed \cite{edin}. The importance of the gluon in diffractive processes
has been anticipated in previous work on the pomeron in QCD \cite{white}.

So far, however, there exists no procedure for calculating
the diffractive structure function in QCD in a systematic way.
It is the goal of this paper to make some progress in this direction.
In view of the phenomenological
success of the results presented in \cite{heb} we shall start from
the production of a quark-antiquark pair as basic process. We are particularly
interested in the non-perturbative mechanism which yields colourless
diffractive final states. As a
first approximation we shall therefore consider the extreme case where
also the partonic gluon in electron-gluon scattering is soft, so that it
cannot
be distinguished from the soft gluons which provide the colour neutralization.
In this limit the diffractive process becomes the production of a
quark-antiquark pair in the colour field of the proton, where the classical
colour field represents a state with a high density of gluons.
We shall work in the proton rest frame where pair production of quarks is
known to be the dominant process \cite{ioffe,hoyer}.

A formalism to treat soft colour interactions in high energy processes
has been developed by Nachtmann \cite{nacht}. So far, it has been applied to
elastic hadron-hadron scattering at high energies \cite{nacht,dosch}.
Similar ideas have been pursued in connection with heavy quark production
in hadron collisions \cite{collins}. In \cite{nacht}
the propagation of quarks in a soft colour field is calculated in an
eikonal approximation which allows to relate quark-quark scattering
amplitudes to Wilson lines. In the following we shall use and further
develop this approach to describe diffractive processes in deep inelastic
scattering. This requires a derivative expansion starting with the eikonal
approximation and a proper treatment of the colour field inside the proton.

The paper is organized as follows. In sect. 2 quark and antiquark
wave functions in a soft colour field are evaluated in a high energy
expansion. Sect. 3 deals with the
pair production cross section, which is obtained in terms of Wilson lines
and their covariant derivatives. In sect. 4 explicit expressions are
derived for inclusive and diffractive structure functions, making some
assumptions about the colour field inside the proton. Sect. 5 contains
a more general discussion of the connection between inclusive
and diffractive structure functions, and in sect. 6 we summarize our results.

\section{High energy expansion for quark wave functions}

The goal of this paper is to calculate the cross section for the production
of a quark-antiquark pair by a virtual photon in an external colour field.
According to the LSZ-formalism, this requires knowledge of the wave functions
of the outgoing quark and antiquark with 4-momenta $l'$ and $l$,
respectively, which are given by
\beq
\bar{\psi}_u(x)=-\int d^4ye^{il'y}\bar{u}(l')i\pds_yS_F(y,x)\quad,\quad
\psi_v(x)=\int d^4yS_F(x,y)i\pdls_yv(l)e^{ily}\label{wf}
\eeq
Here $S_F$ stands for the Feynman
propagator in the colour  background field $G(x)$. Assuming, that the
field $G$ contains no Fourier modes with frequencies of order
$l_0\, ,\,l_0'$ or larger, the Feynman propagator can be replaced
by the advanced propagator (cf. \cite{nacht}). Hence,  these wave functions
can equally well be defined as solutions of the Dirac equation
\beq
\Ds\ \psi_v(x)=0\ \quad ,\quad \bar{\psi}_u(x) \Dls =0\, ,
\eeq
satisfying the boundary conditions
\beq
\psi_v(x)\rightarrow v(l)e^{ilx}\quad ,\quad \bar{\psi}_u(x)\rightarrow
e^{il'x}\bar{u}(l')\quad\mbox{for}\quad x_0\rightarrow \infty\, .
\eeq
Here the covariant derivatives are given by
\beq
D_{\mu} = \partial_{\mu} + i G_{\mu}\quad ,\quad
\Dl\!{}_{\mu} = \pdl\!\!{}_{\mu} - i G_{\mu}\quad ,\quad
G_{\mu}= {1\over 2}\lambda^a G_{\mu}^a\, ,
\eeq
where $G_{\mu}^a$ is the colour field, and the $\lambda^a$ (a=1...8) are
the SU(3) generators.

We are interested in the wave function of an outgoing quark (antiquark) whose
energy is very large compared to the scale on which the colour field varies.
This suggests an expansion of the form,
\bea
\psi_v(x) &=& e^{il\cdot x}\left(V_0 + V_1 + \ldots \right) v(l)\ ,
\label{antiq}\\ \bar{\psi}_u(x) &=& \bar{u}(l')\left(U_0 + U_1 + \ldots
\right) e^{il'\cdot x}\ ,  \label{quark}
\eea
where $V_{n+1}$ ($U_{n+1}$) is suppressed by one power of $l_0$
($l'_0$) with respect to $V_n$ ($U_n$). $v(l)$ and $u(l')$ are
the usual spinors for massless fermions, satisfying
\beq
\ls\ v(l) = 0\ ,\quad \bar{u}(l') \ls' = 0\ .
\eeq
Inserting the expansion (\ref{antiq}) into the Dirac equation, one finds
\beq
i \Ds\ \left( V_0 + V_1 + \ldots \right) v(l) = \ls
\left(V_0 + V_1 + \ldots \right) v(l)\ .
\eeq
This leads to the recursion relations for the matrices $V_n$ ($n>0$),
\bea
l\cdot D\ V_0 &=& 0\ , \label{dv0}\\
\ls V_n \ls\ &=&\ i \Ds\ V_{n-1}\ls\ .\label{rec}
\eea

In order to solve these equations
it is convenient to introduce the projection operators
\beq
P_+ = {1\over 2 l_0} \ls \gamma_0\ ,\quad
P_- = {1\over 2 l_0} \gamma_0 \ls\ =\ 1 - P_+\ .
\eeq
The matrices $V_n$ can now be written as
\beq
V_n = \sum_{i,j=+,-}V_n^{ij}\, ,\qquad\mbox{where}\qquad V_n^{ij}=P_iV_nP_j
\, .
\eeq
Obviously, only $V^{++}$ and $V^{-+}$ contribute in Eq. (\ref{rec}).
Hence we can set $V^{+-}=V^{--}=0$. For the two remaining terms one
obtains from Eq. (\ref{rec}),
\beq
V_n^{-+} = {i\over 2 l_0} \gamma_0 \Ds\ V_{n-1}P_+\ ,\quad
l\cdot D \ V_{n}^{++}\ =\ -{1\over 2} \ls \Ds\ V^{-+}_{n}\ .
\eeq
Inverting the differential operator $l\cdot D$ with appropriate boundary
conditions the solution of these equations can be written in the compact form
\beq\label{recsol}
V_n\ =\ \left({i\over 2 l\cdot D}\Ds P_+ \Ds \right)^n\ P_+\ V_0\ .\label{Vn}
\eeq
Finally, we have to solve the ordinary differential equation
(\ref{dv0}) for $V_0$. The result is the well known non-abelian
phase factor
\beq\label{phase}
V_0(x,l;G)\ =\ P\ \exp\left({i\over 2}
\int_{x_-}^{\infty}dx_-'\ G_+(x_-',x_+,\vec{x}_{\bot})\right)\ .
\eeq
Here $P$ denotes path ordering, and $x^{\mu}$ is decomposed with respect to
$l^{\mu}$,
\beq\label{lcv}
x_-= {1\over l_0}(l_0 x^0 + \vec{x}\cdot\vec{l})\ ,\
x_+= {1\over l_0}(l_0 x^0 - \vec{x}\cdot\vec{l})\ ,\
\vec{x}_{\bot}\cdot \vec{l} = 0\ .
\eeq
The convention for plus- and minus-components is chosen in a way that
stresses their similarity to the light-cone coordinates used later on. Note,
however, that they are not identical, since we will define the $x^3$-axis
anti-parallel to the photon momentum below (cf. Fig.~1).

Eqs. (\ref{antiq}) and (\ref{recsol})-(\ref{lcv})  give the wave
function of an antiquark propagating in a colour field $G_{\mu}^a$,
which at large times approaches the wave function of a free antiquark
with momentum $l$. The wave function is given as a power series in
$D/l_0$ where $D$ stands for a covariant derivative.

The wave function of an outgoing quark can be obtained in a completely
analogous fashion. The result reads explicitly
\bea
U_n\ &=&\ U_0\ P_- \left(\Dls P_- \Dls {i\over 2 \Dl\cdot l'}\right)^n\ ,\\
U_0(x,l';G)\ &=&\ \exp\left({i\over 2}
\int_{\infty}^{x_-} dx_-' G_+(x_-',x_+,\vec{x}_{\bot})\right)\ .
\eea
In these expressions projection operator and components of $x^{\mu}$
are defined with respect to the quark momentum $l'$.

In the following we shall have to manipulate expressions involving the
non-abelian phase factors $V_0$ and $U_0$. We therefore list some
useful formulae which can be derived by standard methods. Let $U(z,y)$
be the phase factor, path-ordered along a straight line from $y$ to $z$,
\beq
U(z,y) = P \exp\left(-i\int_y^z dx^{\mu}G_{\mu}\right)\, .
\eeq
Then the covariant derivative with respect to the upper end point is
given by
\beq
D_{\mu}(z) U(z,y)
= - U(z,y) (z-y)^{\nu}\int_0^1 t dt U^{-1}_t G_{\mu\nu}(y+t(z-y)) U_t\, ,
\eeq
where
\beq
U_t = U(y+t(z-y),y)\, .
\eeq
Note, that the covariant derivative in the direction of the straight line
vanishes.

The path-ordered integral around a triangle can be expressed in terms of
the field strength as follows,
\bea\label{stokes}
U_{\Delta}(z,y,x) &=& U(x,z) U(z,y) U(y,x) \nn\\
&=& P \exp{\int_0^1 du\  b^{\mu}\  K_{\mu}(u)}\, ,
\eea
with
\beq\label{stokes2}
K_{\mu}(u) = - a^{\nu}\int_0^1 tdt U^{-1}_{ut} G_{\mu\nu}(x+t(a+ub))U_{ut}\, ,
\quad U_{ut} = U(x+t(a+ub),x)\, ,
\eeq
and $a=y-x$, $b=z-y$. For one infinitesimal distance,
$b=\epsilon \hat{b}$, this reduces to
\beq
U_{\Delta} = 1 - \epsilon\ \hat{b}^{\mu}a^{\nu}\ \int_0^1 t dt
U^{-1}_t G_{\mu\nu}(x+t a) U_t\, .
\eeq
As expected, in this case $U_{\Delta}-1$ is proportional to the enclosed
infinitesimal area.

\section{Pair production in a colour field}

\subsection{Matrix elements}

We are interested in the diffractive and inclusive structure functions
of the proton at large momentum transfer of the electron and at small $x$,
\beq
-q^2 = Q^2 \gg \Lambda^2\quad,\qquad x = {Q^2\over 2m_pq_0}\ll 1\, ,
\eeq
where $\Lambda$ is the QCD scale parameter and $q_0$ is the photon energy in
the proton rest frame. Here, the photon energy is much larger than the QCD
scale $\Lambda$ and, at low $x$, the kinematically required momentum transfer
to the proton is very small.
Hence, it appears reasonable to describe the photon-proton
interaction as pair production in a classical colour field.
In case of the diffractive structure function, we shall
consider final states with invariant mass
\beq
M^2 = (l+l')^2 = O(Q^2)\, ,
\eeq
where $l'$ and $l$ are the 4-momenta of quark and antiquark, respectively
(cf. Fig. 1). Our goal is to calculate the leading term
of the cross section in the limit
$x\ll 1$ and $\Lambda^2/Q^2 \ll 1$.

The S-matrix element for pair production is given by the expression
\beq
S_{fi} = ie\epsilon_\mu\int d^4xe^{-iqx}\bar{\psi}_u\gamma^\mu\psi_v\equiv
ie \epsilon_\mu S^\mu\, .\label{sm}
\eeq
Here $\psi_u$ and $\psi_v$ are the wave functions of the produced quark
and antiquark, respectively. They have been calculated in the previous section
and they contain the dependence on the colour field.
To leading order one has (cf. eqs. (\ref{antiq}),(\ref{quark})),
\beq
S_{(0)ab}^\mu = \left(\int d^4xe^{i\Delta x}\bar{u}(l')
U_0(x)\gamma_\mu V_0(x)v(l)\right)_{ab}\, ,
\eeq
where
\beq
\Delta = l+l'-q\, ,
\eeq
and the indices $ab$ denote the colour of the created state.

For the evaluation of longitudinal and transverse structure functions the
combinations $S_\mu^*S^\mu$ and $S_0^*S^0$ will be needed.
Summing over the spins of quark and antiquark, one easily obtains,
\beq\label{S00}
S_{(0)}^\mu S_{(0)\mu}^* = -8(ll')|\tilde{f}(\Delta)_{ab}|^2
    = -4M^2|\tilde{f}(\Delta)_{ab}|^2\, ,
\eeq
with
\beq\label{ft}
\tilde{f}(\Delta)_{ab}=\int d^4x e^{i\Delta x}f(x)_{ab}=\int d^4xe^{i\Delta x}
(U_0(x)V_0(x))_{ab}\, .
\eeq
Note, that here and below the dependence of the matrices $U$, $V$ and $f$
on $l'$ and $l$, as described in the previous section,
is not shown explicitly. As it is obvious
from Eq. (\ref{S00}), the leading term, which is formally of order $l_0l_0'
\sim (q_0)^2$ in the high energy limit, turns out to be only of order
$(q_0)^0$ due to the contraction of the 4-vectors $l$ and $l'$.
Therefore the formally
next-to-leading and next-to-next-to-leading contributions are competitive and
have to be calculated.

Hence, we start from the high energy expansion of $S^\mu$,
\beq
S^\mu=S_{(0)}^\mu+S_{(1)}^\mu+S_{(2)}^\mu+\cdots\, ,
\eeq
where
\begin{eqnarray}
S_{(1)}^\mu&=&\int d^4xe^{i\Delta x}\bar{u}(l')\left( U_1\gamma^\mu V_0+
U_0\gamma^\mu V_1\right)v(l)\\ \nonumber\\
S_{(2)}^\mu&=&\int d^4xe^{i\Delta x}\bar{u}(l')\left( U_1\gamma^\mu V_1+
U_2\gamma^\mu V_0+U_0\gamma^\mu V_2\right)v(l)\, .
\end{eqnarray}
The first three terms of the corresponding expansion of the squared
S-matrix element,
\beq
S^\mu S_\mu^*=A_0+A_1+A_2+\ldots\label{lc}
\eeq
have to be taken into account. $A_0$ is given by Eq. (\ref{S00}). The
two following terms are
\bea
A_1 &=& 2\mbox{Re}S_{(0)}^\mu S_{(1)\mu}^*\, ,\nn\\
A_2 &=& S_{(1)}^\mu S_{(1)\mu}^*+2\mbox{Re}S_{(0)}^\mu S_{(2)\mu}^*\, .
\eea

$A_1$ can be easily evaluated as follows. Ordinary
Dirac algebra manipulations give
\beq
S^{\mu*}_{(0)}S_{(1)\mu}=-2\tilde{f}(\Delta)_{ab}^*\int d^4xe^{i\Delta x}
\mbox{tr}\left[\ls\ls'(U_0V_1+U_1V_0)_{ab}\right]\, .
\eeq
Using Eq. (\ref{Vn}) for $V_1$ together with the easily derived relation
\beq
\Ds P_+\Ds P_+V_0=\Ds\,{}^2P_+V_0 \, ,
\eeq
and the corresponding expressions for $U_1$ and $U_0$, one obtains
\begin{eqnarray}
\mbox{tr}\left[\ls\ls'(U_0V_1+U_1V_0)\right]&=&\frac{i}{2}U_0\,\, \mbox{tr}
\left[\ls\ls'\left(\frac{1}{lD}\Ds\,{}^2P_++P_-\Dls\,{}^2\frac{1}{l\Dl}
\right)\right]V_0 \nn\\
&=&2iU_0\left(-l'D-l\Dl+M^2\cal{O}(\frac{D}{l})\right)V_0 \nn\\
&=&-2i(l+l')^\mu\partial_\mu(U_0V_0)+M^2\cal{O}(\frac{D}{l})\, .
\end{eqnarray}
Here the colour indices have been dropped for brevity. Note, that the inverse
differential operator has disappeared. After performing a
partial integration the leading order term $\sim (q_0)^0$ reads
\beq
A_1=4(Q^2+M^2)|\tilde{f}(\Delta)_{ab}|^2\, ,
\eeq
which is very similar to the expression (\ref{S00}) for $A_0$.

The calculation of $A_2$ can be carried out using the same technique and the
formulae of the last section. It turns out to be convenient to introduce
the total momentum $L\equiv l+l'$
and the momentum fraction $\alpha$,
\beq
l\equiv\alpha L+a\quad,\quad l'\equiv(1-\alpha)L-a\quad,\quad a_0\equiv 0\, .
\label{alpha}
\eeq
$\alpha$ and the four-vector $a$ are defined by Eq. (\ref{alpha}). From the
relation $\vec{a}^2=\alpha(1-\alpha)M^2$ it is clear that to leading
order in our high energy expansion
\beq\label{apprl}
l\approx\alpha L\quad,\quad l'\approx(1-\alpha)L\, .
\eeq
Since $A_2$ is formally of order $(q_0)^0$ its leading part is already
sufficient for our purposes. Therefore, the approximation eqs. (\ref{apprl})
may be used. A straightforward calculation yields the result
\beq
A_2=4\left[\frac{1-\alpha}{\alpha}\tilde{g}_{R\mu}(\Delta)_{ab}^*
\tilde{g}^\mu_R(\Delta)_{ab}+\frac{\alpha}{1-\alpha}
\tilde{g}_{L\mu}(\Delta)_{ab}^*\tilde{g}^\mu_{L}(\Delta)_{ab}\right]
-8\mbox{Re}\tilde{f}(\Delta)_{ab}^*
\tilde{h}(\Delta)_{ab}\, ,\label{a2}
\eeq
where $\tilde{g}_{R\mu}\, ,\, \tilde{g}_{L\mu}$ and $\tilde{h}$ are the
Fourier transforms of (cf. eq. (\ref{ft}))
\beq
g_{R\mu}(x)_{ab}=(U_0(x)D_\mu V_0(x))_{ab}\quad ,\quad g_{L\mu}(x)_{ab}=
(U_0(x)\Dl\!{}_\mu V_0(x))_{ab}\label{gdef}
\eeq
and
\beq
h(x)_{ab}=(U_0(x)\Dl\!{}_\mu D^\mu V_0(x))_{ab}\, .\label{hdef}
\eeq
Now the leading contribution to $S^\mu S^*_\mu$ is explicitly given by Eq.
(\ref{lc}).

The calculation of $S^0S^*_0$ is much simpler. Here, contrary to
$S^\mu S^*_\mu$, the leading term is of order $(q_0)^2$ and there is
no suppression after a contraction of Lorentz indices.
One finds,
\beq\label{s0m2}
S^0S^*_0=8q_0^2|\tilde{f}(\Delta)_{ab}|^2\alpha(1-\alpha)\, .\label{sl}
\eeq

\subsection{Cross section}

Consider the production of a quark-antiquark pair by a virtual photon
with 3-momentum $\vec{q}$ in the time interval $-T/2 < x_0 < T/2$. The
standard formula for the production cross section reads
\beq
d\sigma(\gamma^*\rightarrow q\bar{q})=\frac{1}{2|\vec{q}|T}
\int dX|S_{fi}|^2\, ,
\eeq
where the phase space element is
\beq
dX=\frac{d^3\vec{l}}{(2\pi)^32l_0}\,\frac{d^3\vec{l}'}{(2\pi)^32l_0'}=
\frac{d\Delta_+d\Delta_-d^2\Delta_\perp d\alpha}{8(2\pi)^5}\ .
\eeq
Here a change of variables has been performed from the 3-momenta of
quark and antiquark to the momentum transfer in light-cone coordinates,
$\Delta_{\pm}=\Delta^0\pm \Delta^3$, and the momentum fraction $\alpha$
(cf. (\ref{alpha})). Furthermore, one angular integration has been carried
out. Without contraction with the photon polarization vector,
one has (cf. eq. (\ref{sm}))
\beq\label{sigmn}
d\sigma^{\mu\nu}=\frac{\pi e^2}{|\vec{q}|}\frac{1}{2\pi T}\int
dX S^{\mu}S^{*\nu}\ ,
\eeq
with
\beq
S^\mu=\int d^4xe^{i\Delta x}F^\mu(x;\Delta,\alpha)\ .
\eeq
Here the function $F$ stands for the expressions $f,g$ and $h$ introduced in
Eqs. (\ref{ft}) and (\ref{gdef}), (\ref{hdef}) above. The dependence on
the quark-antiquark configuration, and therefore on $\Delta$ and $\alpha$,
is explicitly kept.

It is useful to also write the photon 4-momentum $q^{\mu}=
(q_0,\vec{0},-\sqrt{q_0^2+Q^2})$ (cf. Fig. 1) in light-cone coordinates,
\beq
q = (q_+, q_-, \vec{q_{\perp}}) = \left(-{Q^2\over 2q_0},2q_0,\vec{0}
\right)\ .
\eeq
In the proton rest frame the large component is $q_-$. The momentum
transferred by the smoothly varying external field is small, $\Delta = O(
\Lambda)$. Hence, $\Delta_- \ll q_-$ is irrelevant for the quark-antiquark
configuration.
This is also clear from the expression for the opening angle $\theta$, which,
in the configuration with $\vec{l}_{\perp}+\vec{l'}_{\perp}=0$,
reads
\beq\label{angle}
\theta^2 \simeq {M^2 \over q_0^2 \alpha (1-\alpha)}\ ,
\eeq
with the invariant mass $M^2$ given by
\beq
M^2+Q^2=q_+\Delta_-+q_-\Delta_++\Delta^2\simeq 2q_0\Delta_+\, .
\eeq
We choose the coordinate system such that the proton is at the origin, and
$x_\p\equiv x^3$ is the distance between the proton and the point where the
pair is created. Since the quark-antiquark configuration is essentially
independent of $\Delta_-$, the same can be expected to hold for
$F(x;\Delta,\alpha)$. Hence, we have
\bea\label{fapprox}
F(x;\Delta,\alpha) &\equiv& F(x_+,x_\p,x_\perp;\Delta_+,\Delta_-,\Delta_\perp,
\alpha) \nn\\
&\simeq& F(x_+,x_\p,x_\perp;\Delta_+,\Delta_\perp,\alpha)\, .
\eea
Using this property of the function $F$, we can carry out the integrations
over $\Delta_-$, $x_+$ and $x'_+$ in Eq. (\ref{sigmn}). Dropping Lorentz
indices for brevity, one obtains
\beq
d\sigma\simeq{\pi e^2\over |\vec{q}|}{1\over 4(2\pi)^5T}
\int d\Delta_+d^2\Delta_\perp d\alpha\, I(\Delta,\alpha)\, ,
\eeq
where
\bea
I(\Delta,\alpha)&&\!\!\!\!\!\!\!\!=\!\int d^4x\int d^4x'\delta(x_+-x'_+)
\;e^{i(\frac{x_-\Delta_+}{2}-x_\perp\Delta_\perp)}\;
e^{-i(\frac{x'_-\Delta_+}{2}-x'_\perp\Delta_\perp)}\;
F(x;\Delta,\alpha)F^*(x';\Delta,\alpha) \nn
\\
\nn\\&&\!\!\!\!\!\!\!\!=\!\int d^2x_\perp dx_+ \!\!\!\!
\int\limits_{-T/2+x_+}^{T/2+x_+}\!\!\!\!dx_\p\,\,\int d^2x'_\perp dx'_+\!\!\!
\!\int\limits_{-T/2+x'_+}^{T/2+x'_+}\!\!\!\!dx'_\p\,\,\delta(x_+-x_+')\nn
\\
\nn\\&&\qquad\qquad\qquad
e^{-i(x_\p\Delta_+ +x_\perp\Delta_\perp)}\;
e^{i(x'_\p\Delta_+ +x'_\perp\Delta_\perp)}\;
F(x;\Delta,\alpha)F^*(x';\Delta,\alpha)\nn
\\
\nn\\&&\!\!\!\!\!\!\!\!=\!\!\!\int\limits_{-\infty}^{+\infty}dx_+\left|
\,\int\limits_{-T/2+x_+}^{T/2+x_+}\!\!\!\!dx_\p\int d^2x_\perp
\;e^{-i(x_\p\Delta_+ +x_\perp\Delta_\perp)}\;F(x_+,x_\p,x_\perp;\Delta_+,
\Delta_\perp,\alpha)\right|^2\nn
\\
\nn\\&&\!\!\!\!\!\!\!\!=\!\!\!\int\limits_{-T/2}^{+T/2}dx_+\left|
\,\int\limits_{-\infty}^{+\infty}\!\!dx_\p\int d^2x_\perp
\;e^{-i(x_\p\Delta_+ +x_\perp\Delta_\perp)}\;F(x_+,x_\p,x_\perp;\Delta_+,
\Delta_\perp,\alpha)\right|^2\, .\label{fi}
\eea
The last equality holds if only a finite range $L \ll T$ contributes
to the integration over $x_\p$. Technically, this can be
realized by introducing some smooth suppression of $F$ for $x_\p\to\infty$,
which is removed after the limit $T\to\infty$ has been taken.

The cross section now reads
\beq\label{fplus}
d\sigma\simeq\,\frac{1}{T}\!\!\!\!\int\limits_{-T/2}^{T/2}\!\!\!\!dx_+\,
\left\{\frac{\pi e^2}{|\vec{q}|}\frac{1}{4(2\pi)^5}\int d\Delta_+ d^2
\Delta_\perp d\alpha\left|\int d^3x\; e^{-i(x_\p\Delta_++x_\perp\Delta_\perp
)}F(x_+,\vec{x};\Delta,\alpha)\right|^2\right\} \, .
\eeq
This expression is an average over all times $x_+$ at which the proton is
tested by the quark-antiquark pair. For a smoothly varying colour field
the $x_+$-dependence should not matter and the corresponding integral
is trivial. This yields the final result
\beq
d\sigma\simeq\frac{\pi e^2}{|\vec{q}|}\frac{1}{4(2\pi)^5}\int d\Delta_+ d^2
\Delta_\perp d\alpha\left|\int d^3x e^{-i(x_\p\Delta_++x_\perp
\Delta_\perp)}F(\vec{x};\Delta,\alpha)\right|^2 \, ,\label{cs}
\eeq
which is similar to the cross section for pair production
in a static colour field.
The only difference lies in the Fourier transform of the function $F$, where
$\Delta_+$ occurs instead of $\Delta_\p$.

The expressions for $S^\mu S^*_\mu$ and $S^0S^*_0$ obtained in Sect. 3.1
contain the 4-dimensional Fourier transforms $\tilde{f}$, $\tilde{g}$
and $\tilde{h}$. They have to be replaced by the 3-dimensional Fourier
transforms appearing in Eq. (\ref{cs}), when used in expressions for
cross sections in the next section.

\section{Inclusive and diffractive structure functions}
\subsection{Relations between different formfactors}

Before discussing the actual functional form of $f,\, g_R,\, g_L$ and $h$,
introduced in Sect. 3.1, we shall derive some relations between
terms involving $g_R$ and $g_L$. For brevity, colour indices will be dropped
throughout this subsection.

As we shall see in the following subsection, the Fourier integrals
$\tilde{f}$, $\tilde{g}_{L,R}$ and $\tilde{h}$ receive negligible
contributions from the region where the spatial point $x$ lies inside
the proton. Therefore, one may use the approximation
\beq
g_{R\mu}(x) \simeq U_0(x)\partial_\mu V_0(x)\quad,\quad
g_{L\mu}(x) \simeq U_0(x)\pdl\!\!{}_\mu V_0(x)\, .
\eeq
Since $l$ is a light-like vector with negligible transverse components,
and $l^\mu g_\mu = 0$, Eq. (\ref{a2}) can be written as
\beq
A_2=-4\,\frac{1-\alpha}{\alpha}\sum_{i=1}^{2}|\tilde{g}_{Ri}(\Delta)|^2-4\,
\frac{\alpha}{1-\alpha}\sum_{i=1}^{2}|\tilde{g}_{Li}(\Delta)|^2-8\mbox{Re}
\tilde{f}(\Delta)^*\tilde{h}(\Delta)\, .\label{sum}
\eeq
In the following, we mean by $\tilde{g}_R$, $\tilde{g}_L$,
$\tilde{f}$ and $\tilde{h}$
the 3-dimensional Fourier transforms in the sense of Eq. (\ref{cs}) which
occur in the cross sections. Correspondingly, $A_i$, $i=0,1,2$, are the
expressions defined in Sect. 3.1 in terms of the 3-dimensional Fourier
transforms.

It is convenient to further specify the coordinate system. In addition
to choosing the $x^3$-axis anti-parallel to the photon momentum,
we assume the plane spanned by $\vec{l}$ and $\vec{l'}$
to be orthogonal to the $x^1$-axis, neglecting the small transverse
momentum transfer. Defining $\theta_1,\,\theta_2$ and $\theta$ to be
the angles between $\vec{l'}$
and $\vec{q}$, between $\vec{q}$ and $\vec{l}$ and between $\vec{l'}$ and
$\vec{l}$, respectively, and writing $-i(x_\p\Delta_++x_\perp\Delta_\perp)=ix
\Delta$ for brevity, one has
\bea\label{g2}
\hspace*{-.5cm}i\Delta_+\int e^{ix\Delta}U_0V_0&=&
\int e^{ix\Delta}\partial_3(U_0V_0)\nn\\
&=&\int e^{ix\Delta}\left(\theta_1U_0
\partial_2\!\!\!\!\!\raisebox{1.5ex}{$\leftarrow$}
V_0-\theta_2U_0\partial_2V_0
\right) \nn\\
&=&\int e^{ix\Delta}\left(-i\Delta_2\theta_1U_0V_0-
\theta_1U_0\partial_2V_0-\theta_2U_0\partial_2V_0\right)\nn\\
&=&-\int e^{ix\Delta}
\left(i\Delta_2\theta_1U_0V_0+\theta U_0\partial_2V_0\right)\, .
\eea
In order to obtain the second equality, one has to observe that in
our coordinate system, moving e.g. the quark-line, associated with $U_0$,
by some small amount $\epsilon$ in $x^3$-direction is equivalent to
moving it by the amount $\theta_1\epsilon$ in $x^2$-direction.
{}From Eq. (\ref{g2}) we obtain the relation
\beq
\left|\int e^{ix\Delta}U_0\partial_2V_0\right|=\frac{\Delta_+}{\theta}
|\tilde{f}|+\frac{\Delta_2\theta_1}{\theta}|\tilde{f}|\, .\label{d20}
\eeq

The first term on the r.h.s. of Eq. (\ref{d20}) gives a contribution of
the same order of magnitude as the formally leading terms $A_0$ and $A_1$.
However, the contribution proportional to the transverse
momentum transfer $\Delta_2$ can be neglected. Obviously, as long as $\alpha$
and $1-\alpha$ are $O(1)$,
it is suppressed with respect to the leading contributions by
a factor $|\Delta_2|/Q \sim \Lambda/Q$. This suppression is not so
obvious if $\alpha\ll 1$, since then the prefactor $(1-\alpha)/\alpha$
in Eq. (\ref{sum}) becomes large. Yet this enhancement is completely
compensated by the factor $\theta_1/\theta$, since in this limit $\theta_1/
\theta=\theta_1/(\theta_1+\theta_2)\approx\theta_1/\theta_2\approx\alpha/(1-
\alpha)$. These relations can be read off from the vector parallelogram
corresponding to $\vec{l}+\vec{l'}=\vec{q}$. Hence, we obtain for the
leading contribution
\beq
|\tilde{g}_{R2}(\Delta)| \simeq \left|\int e^{ix\Delta}U_0\partial_2V_0\right|
\simeq \frac{\Delta_+}{\theta}|\tilde{f}|\label{d2}\, .
\eeq

The term in Eq. (\ref{sum}) involving $\tilde{g}_{R1}$ can not be directly
related to $\tilde{f}$. However, for very small and very large values of
$x_\p$, one expects
\beq
\left|\int e^{ix\Delta}U_0\partial_1V_0\right|\simeq
\left|\int e^{ix\Delta}U_0\partial_2V_0\right|\label{ffp}\, .
\eeq
For sufficiently small $x_\p$ one has $U_0 \simeq 1$, which restores
the rotational invariance in the transverse plane.
For sufficiently large $x_\p$ only the quark or the antiquark trajectory
penetrates the proton field. Hence, either $U_0=1$ or $V_0=1$,
thus again restoring rotational invariance.

The above considerations suggest to replace $\tilde{g}_{R1}$ by
a function $\tilde{f}'$, defined in analogy to Eq. (\ref{d2}),
\beq
|\tilde{g}_{R1}(\Delta)|=\left|\int e^{ix\Delta}U_0\partial_1V_0\right|=
\frac{\Delta_+}{\theta}|\tilde{f'}|\, .\label{d1}
\eeq
The functions $\tilde{f}$ and $\tilde{f}'$ are then expected to have
a similar asymptotic behaviour.

The arguments used above to rewrite the terms involving $\tilde{g}_R$ also
apply to the terms involving $\tilde{g}_L$.

Combining the results of Sect. 3.1 for $A_0$ and $A_1$ with Eqs.
(\ref{angle}), (\ref{sum}), (\ref{d2}) and (\ref{d1}), we obtain for the
leading contribution to $S^{\mu}S^*_{\mu}$,
\beq
A_0+A_1+A_2=Q^2\left[4|\tilde{f}|^2-\frac{1-2\alpha(1-\alpha)}
{\beta(1-\beta)}\left(|\tilde{f}|^2+|\tilde{f}'|^2\right)\right]
-8\mbox{Re}\tilde{f}^*\tilde{h}\, ,
\eeq
where
\beq
\beta=\frac{Q^2}{Q^2+M^2}
\eeq
is the parameter conventionally introduced in diffractive deep-inelastic
scattering. The corresponding expression for $S^0S^*_0$ is given in
Eq. (\ref{s0m2}). Inserting these expressions in Eq. (\ref{cs}) yields
the $\gamma^*p$ cross sections
\bea
d\sigma^\mu{}_\mu &=&\frac{\pi e^2}{|\vec{q}|}\frac{1}{4(2\pi)^5}\!
\int d^2\Delta_\perp d\Delta_+d\alpha \nn\\
&&\qquad\qquad\times \left\{Q^2\left[4|\tilde{f}|^2\!
-\!\frac{1-2\alpha(1-\alpha)}{\beta(1-\beta)}\left(|\tilde{f}|^2
+|\tilde{f}'|^2\right)\right]  -\!8\mbox{Re}\tilde{f}^*
\tilde{h}\right\}\, ,\label{smm}\\
d\sigma_{00} &=&\frac{\pi e^2}{|\vec{q}|}\frac{2}{(2\pi)^5}\!
\int d^2\Delta_\perp d\Delta_+d\alpha
q_0^2|\tilde{f}|^2\alpha(1-\alpha)\, .\label{s00}
\eea
{}From these two cross sections one can obtain transverse and longitudinal
structure functions in the usual way.

\subsection{Averages over the proton colour field}

All the information on the photon-proton interaction is contained in
the functions $\tilde{f}(\Delta)_{ab},\,\tilde{f}'(\Delta)_{ab}$ and
$\tilde{h}(\Delta)_{ab}$, which occur in Eqs. (\ref{smm})
and (\ref{s00}). As discussed in Sect. 3.2,
the 3-dimensional Fourier transforms with respect to $x_\p$ and $x_\perp$,
as defined in Eq. (\ref{cs}), are needed for
the cross sections and the related structure functions. In this section
several general features of these functions will be discussed, which
will allow us to evaluate the inclusive and diffractive structure functions
in terms of several unknown constants. Our main assumptions are that
the field strength $G_{\mu\nu}(x)$ vanishes outside a region
of size $\sim 1/\Lambda$, and that it varies smoothly on a scale $\Lambda$.

Consider first the dependence on the transverse coordinates $x_\perp$. For
$f(x)_{ab}$, $a\neq b$, to be non-zero, at least one of
the two fermion lines of Fig. 1 has to pass through the region with non-zero
field, which has the transverse size $\sim 1/\Lambda$. Hence, integration over
the transverse coordinates can be expected to yield
\beq
\int d^2x_\perp e^{-ix_\perp\Delta_\perp}
f(x_\p,x_\perp;\Delta_+,\Delta_\perp,\alpha)_{ab}
\simeq \frac{\pi}{\Lambda^2}\exp(-\frac{\Delta_\perp^2}{4\Lambda^2})
f_\p(x_\p;\Delta_+,\alpha)_{ab}\, .
\label{fpar}
\eeq
This relation becomes exact for $f(x;\Delta,\alpha)=f_\p(x_\p;\Delta_+,\alpha)
\exp(-x_\perp^2\Lambda^2)$, but its qualitative features can be expected
to hold in general. The case $a=b$ can be treated completely analogous, after
replacing $f_{aa}$ by $f_{aa}-1$. This is possible since the constant does
not contribute to the Fourier transform for $\Delta_+\neq 0$.

According to eqs. (\ref{stokes}) and (\ref{stokes2}),
given at the end of Sect. 2, the function
$f_{ab}$ integrates the colour field strength in the double shaded
area of Fig. 1. Outside the shaded area the Wilson lines of quark
and antiquark may be connected by a space-like Wilson line in the
quark-antiquark plane, yielding a Wilson triangle.
The light-like vector $a^\mu$ essentially points along
the light-cone axis $x_+ = x^0 +x^3$ = const., and the space-like vector
$b^\mu$ may be chosen orthogonal to the longitudinal axis.
The component of the field strength tensor, which is integrated over the
double shaded area, reads
\beq
G_{+,2\ ab}\ = E_{2\ ab}\ - \ B_{1\ ab}\, ,
\eeq
where we have chosen the 1-axis to be perpendicular to the quark-antiquark
plane, $G_{0i}=E_i$ and $G_{ij}=-\epsilon_{ijk}B_k$.

For small areas the integral $f_{ab}$ is proportional to the area.
Since we are considering invariant masses $M^2 = O(Q^2)$, the opening
angle $\theta$ is always small. Hence, $f_\p$ rises linearly in $x_\p$
and $\theta$ in the range $0<x_\p\ll 1/\theta\Lambda$ ($a\neq b$),
\beq
f_\p(x_\p; \Delta_+,\alpha)_{ab}\simeq
C_{ab}\, x_\p\,\theta\Lambda\, .\label{C}
\eeq
For $x_\p \sim 1/\theta\Lambda$, the area reaches the size of the proton
area $\sim 1/\Lambda^2$. Since the average field strength is
$\sim \Lambda^2$, one has $f_\p (1/\theta\Lambda) = O(1)$, and therefore
$C=O(1)$. Also, for $x_\p \gg 1/\theta\Lambda$, $f_\p$ must be bounded by
a constant $O(1)$. Eq. (\ref{C}) then implies for the Fourier transform
in the range $\Delta_+ \gg \theta\Lambda$,
\beq
\tilde{f}_\p(\Delta_+,\alpha)_{ab}\simeq \frac{C_{ab}}{\Delta_+}\,
\frac{\theta\Lambda}{\Delta_+}\, .\label{cv}
\eeq
This behaviour is intuitively clear since high frequency
modes are only present due to the onset of the linear rise at $x_\p=0$.
In the range $\Delta_+ \ll \theta\Lambda$, the Fourier transform
$\tilde{f}_\p$ is bounded by $\sim 1/\Delta_+$ on dimensional grounds.

As suggested by Eq. (\ref{cv}) it proves convenient to introduce the variable
\beq\label{y}
y^2\equiv\left(\frac{\theta\Lambda}{\Delta_+}\right)^2
=\frac{z^2}{\alpha(1-\alpha)}\, ,\\
\eeq
with
\beq\label{z}
 z^2=4\beta(1-\beta)\frac{\Lambda^2}{Q^2}\, .
\eeq
Note, that up to corrections of relative order $\theta$, the $x_\p$-dependence
of $f_\p(x_\p;\Delta_+,\alpha)$ is a dependence on the product
$\theta x_\p$. This is true, since after the $x_\perp$-integration
the transverse distance at which the two quarks penetrate the proton, i.e.
$\theta x_\p$, is the only relevant parameter.

As a result, the Fourier transform $\tilde{f}_\p$ can only depend in a
non-trivial way on the dimensionless variable $y$. Therefore we introduce a
dimensionless function $\bar{f}$, requiring
\beq\label{fy}
\tilde{f}_\p(\Delta_+,\alpha)_{ab} \simeq {1\over \Delta_+}
\bar{f}(y)_{ab}\, ,
\eeq
where $\bar{f}(y)$ has the properties (cf. (\ref{C}),(\ref{cv})),
\beq
\bar{f}(0)_{ab}=0\, ,\quad
\frac{\partial}{\partial y}\bar{f}(y)_{ab}\Bigg|_{y=0} = C_{ab} = O(1)\, ,
\quad\mid \bar{f}(y)_{ab}\mid\, <\, O(1)\quad\mbox{as}\quad y\to\infty\, .
\eeq
In analogy to the function $\bar{f}(y)_{ab}$,
one can define functions $\bar{f}'(y)_{ab}$ and
$\bar{h}(y)_{ab}$ starting from the functions $\tilde{f}'(\Delta)_{ab}$
and $\tilde{h}(\Delta)_{ab}$ which occurred in the previous subsection.

We are know ready to evaluate the wanted structure functions. In the proton
rest frame they are related to the cross sections by the well known relations
\bea
F_\Sigma &=& -\frac{Q^2}{2\pi e^2}\ \sigma^\mu{}_\mu\quad,\label{f2s}\\
F_L &=& \frac{Q^4}{\pi e^2q_0^2}\ \sigma_{00}\quad, \label{f2l}\\
F_2 &=& F_\Sigma+\frac{3}{2}F_L\quad . \label{f2}
\eea
Since the diffractive structure functions are usually defined in terms of
$x,\, Q^2$ and $\xi\equiv x/\beta=\Delta_+/m_p$, we change variables
from $\Delta_+$ to $\xi$ and also from $\alpha$ to $y$.
Performing the $\Delta_\perp$-integration in Eqs. (\ref{smm}) and (\ref{s00})
and dropping colour-indices for the rest of this subsection, one obtains
\bea
d F_\Sigma&=&\frac{\beta d\xi}{4\pi^2\xi}\int\limits_{2z}^\infty\frac{dy}
{y^2\sqrt{y^2-4z^2}}\times\label{FS}\\ \nonumber\\&&
\times\left[(1-\frac{2z^2}{y^2})(|\bar{f}|^2+|\bar{f}'|^2)-4\beta(1-\beta)
|\bar{f}|^2+\frac{8\beta(1-\beta)}{Q^2}\mbox{Re}\bar{f}^*\bar{h}\right]
\, ,\nn\\
&& \nn\\
dF_L&=&\frac{4\beta^2(1-\beta)d\xi}{\pi^2\xi}\int\limits_{2z}^\infty
\frac{z^2\, dy}{y^4\sqrt{y^2-4z^2}}|\bar{f}|^2\, .\label{FL}
\eea
Finally, the integral over $\xi$ has to be carried out.

The behaviour of the structure functions at large $Q^2$ can be found by
calculating the $y$-integrals in the above two formulas in the limit
$z\to 0$ (cf. (\ref{z})). As an illustration we discuss the integral
\beq
J=\int\limits_{2z}^\infty\frac{dy}{y^2\sqrt{y^2-4z^2}}|\bar{f}|^2\, .
\eeq
Splitting $J$ into the two contributions
\bea
J &=& J_1+J_2 \, , \nn\\
J_1 &=& \int\limits_{2z}^\infty\frac{dy}{\sqrt{y^2-4z^2}}\,\frac{1}
{(1+y^2)}|C|^2\, , \nn\\
J_2 &=& \int\limits_{2z}^\infty\frac{dy}{y^2\sqrt{y^2-4z^2}}
\left(|\bar{f}|^2-\frac{y^2}{1+y^2}|C|^2\right)\, ,
\eea
the limit $z\to 0$ can be performed for $J_2$, while the first part, which
can be calculated explicitly, diverges logarithmically in this limit.
Therefore, dropping all contributions vanishing for $z\to 0$,
\beq
J=|C|^2\ln\frac{1}{z}+{\cal F}[\bar{f}]\, ,
\eeq
where the functional ${\cal F}$ is defined by
\beq
{\cal F}[\bar{f}]=\int\limits_0^\infty\frac{dy}{y^3}\left(
|\bar{f}|^2-\frac{y^2}{1+y^2}|C|^2\right)\, .
\eeq

The other contributions in Eqs. (\ref{FS}) and (\ref{FL}) can be evaluated
in the same way. Since the function $h(x)$ contains two covariant
derivatives, and therefore two gauge potentials, one expects that
$\bar{h}$ is bounded by const.$\times\Lambda^2$. In Eq. (\ref{smm}) the
term involving $\bar{h}$ is suppressed by $1/Q^2$ and does therefore
not contribute in the limit $z\to 0$. Using the relation
$|C_{ab}'| =|C_{ab}|$ for the derivative of $\bar{f}'$ at $y=0$,
which follows from the validity of Eq. (\ref{ffp}) in the region
of small $x_\p$, we finally obtain
\bea
dF_\Sigma&=&\frac{\beta d\xi}{4\pi^2\xi}\left[\left\{\left(2-4\beta(1-\beta)
\right)\ln\frac{1}{z}-1\right\}|C|^2+\left(1-4\beta(1-\beta)
\right){\cal F}[\bar{f}]+{\cal F}[\bar{f}']\right]\label{fs}\\ \nonumber\\
dF_L&=&\frac{\beta^2(1-\beta)d\xi}{\pi^2\xi}|C|^2\, .\label{fl}
\end{eqnarray}

To conclude this subsection, let us recall our main assumptions which led to
this result. We have assumed that the proton colour field is confined to a
region of size $\sim 1/\Lambda$, that it varies smoothly on a scale $\Lambda$,
and that $\tilde{f}$ and $\tilde{f}'$ increase linearly with the probed area
of the proton for small longitudinal separation.

\subsection{Results for inclusive and diffractive structure functions}

The inclusive structure functions can now be obtained from Eqs. (\ref{fs}),
(\ref{fl}) by substituting $d\xi/\xi=-d\beta/\beta$, performing the
$\beta$-integration in the kinematical limits $x<\beta<1$, and summing over
the colours in the final state. Dropping contributions suppressed by $x$ and
introducing the two constants
\beq
C_1=\sum_{a,b}|C_{ab}|^2\quad,\quad C_2=\sum_{a,b}\left\{
\frac{13}{6}|C_{ab}|^2+\frac{1}{2}{\cal F}[\bar{f}_{ab}]+
\frac{3}{2}{\cal F}[\bar{f}'_{ab}]\right\}\, ,
\eeq
which are expected to be $O(1)$, one obtains
\bea
F_2(x,Q^2)&=&\frac{1}{6\pi^2}\left(C_1\ln\frac{Q^2}{4\Lambda^2}+C_2\right)
\, ,\label{f20}\\
F_L(x,Q^2)&=&\frac{1}{6\pi^2}C_1\, .\label{fl0}
\eea
The leading contribution as $x \to 0$ is non-singular. Since our
semiclassical approach in its present form does not contain any
QCD-radiation effects, this is not surprising.
As expected, the semiclassical approximation is meaningful
in the regime of very high parton densities, and it therefore leads to
a result for which the limit of $x\to 0$ exists.
Note, that Eqs. (\ref{f20}) and (\ref{fl0}) are very similar to
results obtained for pair production in an external field in quantum
electrodynamics \cite{soper}.

In the range of $x$ presently probed at HERA, an increase of the structure
function with decreasing $x$ is theoretically expected \cite{dglap, bfkl}
and experimentally observed \cite{zeusf2,h1f2}. In order to obtain
a realistic description of structure functions in this range of $x$,
radiation effects have to be taken into account, which may be possible
following the approach of Lipatov \cite{lip} and Balitsky \cite{bal}.

The $\ln{Q^2}$ term in $F_2$ is due to the integration
over all possible configurations of quark-antiquark pairs, from
symmetric pairs with $\alpha\approx 1/2$
to extremely  asymmetric pairs with
$\alpha\approx \Lambda^2/Q^2$ (or $1-\alpha\approx\Lambda^2/Q^2$) very small.
In the latter case one particle
carries most of the momentum of the photon.
The absence of such a logarithm in $F_L$ can be traced back to the
factor $\alpha(1-\alpha)$ in Eq. (\ref{sl}), which
suppresses the contribution of asymmetric pairs. A similar observation
has been made in \cite{brodsky} in connection with the wave function
of the longitudinal photon.

It is also very instructive to compare this with the origin of the
$\ln{Q^2}$ term in the usual perturbative treatment in the infinite
momentum frame. In the contribution to $F_2$ from photon-gluon fusion
the appearance of the logarithm is due to configurations where the
produced quark or antiquark is almost collinear with the gluon. In the
proton rest frame this corresponds to a configuration where one of the
produced particles is relatively soft.

Diffractive events are expected to occur whenever the quark-antiquark pair is
created in a colour-singlet state. In this case the proton remnant is also in
a singlet and no colour flow occurs between the remnant, most probably still a
proton, and the created fast moving pair.  Hence, in the hadronic energy
flow a large gap in rapidity is expected.

In order to obtain the diffractive structure functions, one has to project
on colour-singlet final states in Eqs. (\ref{fs}) and (\ref{fl}).
This amounts to the replacement
\beq
\bar{f}_{ab}\,\to\,\frac{1}{\sqrt{3}}\mbox{tr}\bar{f}=\frac{1}{\sqrt{3}}
\sum_a \bar{f}_{aa}\, ,
\eeq
and the corresponding substitutions for $\bar{f}'$ and $\bar{h}$.

It is now important to observe that, in contrast to $f_\p(x_\p)_{ab}$ the
function tr$f_\p(x_\p)$ does not exhibit the linear rise at small $x_\p$,
as considered in Eq. (\ref{C}). This is easily understood remembering that
the contribution to $f(x)_{ab}$ linear in $\theta$ is also linear in the
field strength $G^{\mu\nu}{}_{ab}$, which is traceless (cf. Sect. 2).
Therefore, tr$f_\p(x_\p)$ is rising like $(x_\p)^2$ at small $x_\p$.
For the Fourier transform this means, that tr$\bar{f}(y)\sim y^2$
for $y\ll 1$, i.e. tr$C=0$.

Since the diffractive structure functions are defined by cross-sections
differential in $\xi$, no integration needs to be performed in Eqs.
(\ref{fs}) and (\ref{fl}). $F_2^D$ and $F_L^D$ are then
obtained by dropping terms proportional to $|C|^2$ and
by substituting for $\bar{f}_{ab}\, ,\, \bar{f}_{ab}'$ the
correctly normalized traces. This yields
\bea
F_2^D(x,Q^2,\xi)&=&\frac{\beta}{4\pi^2\xi}\left[\left\{1-4\beta(1-\beta)
\right\}C_3+C_4\right]\, ,\\
F_L^D(x,Q^2,\xi)&=&0\quad ,
\eea
where
\beq
C_3={\cal F}[{1\over \sqrt{3}}\mbox{tr}\bar{f}]\quad, \quad
C_4={\cal F}[{1\over \sqrt{3}}\mbox{tr}\bar{f}']\, .
\eeq
have been introduced. Note, that the vanishing of tr$C$
simplifies the form of the functional ${\cal F}$, resulting in
\beq
C_3={1\over 3}\int\limits_0^\infty\frac{dy}{y^3}|\mbox{tr}\bar{f}|^2\quad,
\quad C_4={1\over 3}\int\limits_0^\infty\frac{dy}{y^3}|\mbox{tr}
\bar{f}'|^2\, .
\eeq
Due to the slower rise of tr$\bar{f}(y)$ at small $y$, the contribution
of symmetric quark-antiquark pairs is suppressed and only asymmetric
pairs are relevant for the leading twist contribution to the diffractive
structure function. This explains the absence of a term
$\sim \ln(Q^2/\Lambda^2)$ in $F_2^D$ and the vanishing of $F_L^D$.

In summary, we arrive at a clear picture of the final states in ordinary
deep inelastic and diffractive deep inelastic events. Symmetric and
asymmetric pairs contribute to $F_2$. This leads to a contribution growing
logarithmically with $Q^2$. In the diffractive case, the contribution
from symmetric pairs  is suppressed, implying
the absence of a term $\sim \ln(Q^2/\Lambda^2)$. For both,
the inclusive and the diffractive longitudinal structure function,
the contribution of asymmetric pairs is suppressed.
As a result, $F_L$ contains no term $\sim \ln(Q^2/\Lambda^2)$,
and $F_L^D$ is suppressed by $1/Q^2$, i.e. of higher twist.

\section{On the relation between $F_2$ and $F_2^D$}

The above results are similar to those obtained from the simple partonic
picture of \cite{heb}, as far as the relation between the inclusive
structure function $F_2$ and the diffractive structure function $F_2^D$ is
concerned. In both models $F_2^D$ is related to $F_2$ by some constant
connected with the colour-singlet requirement for the produced
quark-antiquark pair, and in both cases the slope in $x$ of $F_2$
is larger by one unit than the slope of $F_2^D$ in $\xi$.

This relation between the slopes of the two structure functions has first been
proposed in \cite{buch} based on a picture of diffraction as
scattering on wee parton lumps inside the proton. In the present
section we will argue, that there is a rather general class of models for
diffractive deep inelastic scattering,
which should reproduce the above relation.

Since in small-$x$ events the kinematically required momentum transfer to the
proton-target is relatively small, it is natural to think of these events
in terms of a virtual photon cross-section $\sigma_T(\gamma^*p\to p'X)$.
Here $p'$ is the proton remnant, which, in this picture, can be separated
from the produced massive state $X$ before hadronization. Assume
that, with $X$ being in a colour-singlet state, no hadronic activity develops
between $X$ and $p'$, and a rapidity gap event occurs. Now, neglecting the
longitudinal contribution, the structure functions read
\begin{eqnarray}
F_2(x,Q^2)&=&\frac{Q^2}{\pi e^2}\int\limits_x^1\frac{d\sigma_T(\gamma^*p\to
p'X)}{d\xi}d\xi\nonumber\\ \\
F_2^D(x,Q^2,\xi)&=&\frac{Q^2}{\pi e^2}\cdot\frac{d\sigma_T(\gamma^*p\to p'X)}
{d\xi}\cdot{\cal P}\, ,\nonumber
\end{eqnarray}
where the probability for the proton remnant to be in a colour-singlet state
is given by the factor ${\cal P}$. This results in the relation
\beq
F_2(x,Q^2)=\int\limits_x^1d\xi F_2^D(x,Q^2,\xi)\cdot{\cal P}^{-1}=x
\int\limits_x^1d\beta\,\beta^{-2}F_2^D(x,Q^2,\xi)\cdot{\cal P}^{-1}\, .
\eeq
We now assume that the $\xi$-dependence of $F_2^D$ factorizes,
\beq
F_2^D(x,Q^2,\xi)=\xi^{-n}\hat{F}_2^D(\beta,Q^2)\, ,
\eeq
with some number $n>1$. This is consistent with data in a
wide region of $\xi$ (cf. \cite{zeus,h1}). Dropping terms suppressed in the
limit $x\to 0$, the relation between diffractive and inclusive structure
function takes the form
\beq
F_2(x,Q^2)=x^{1-n}\int\limits_0^1d\beta\,\beta^{n-2}\hat{F}_2^D(\beta,Q^2)
\cdot{\cal P}^{-1}\, .\label{f2f2d}
\eeq
At small $x$ the photon energy $q_0$ in the proton rest frame is much larger
than any other scale in the problem. Therefore it is natural to consider
the limit $q_0\to\infty$ and to try to understand the presence of a non-zero
probability factor ${\cal P}$ in this limit. Assuming that this small-$x$
limit for ${\cal P}$ has already been reached in the present measurements,
${\cal P}$ is a function of $\beta$ and $Q^2$ only.
In this case Eq. (\ref{f2f2d})
implies that the $x$-slope of $F_2$ and the $\xi$-slope $F_2^D$ differ
by one unit. It is interesting to observe that the result of this rather hand
waving argument appears to agree well with the data on $F_2^D$
\cite{zeus,h1} and $F_2$ \cite{zeusf2,h1f2} at small $x$.

\section{Conclusions}
A semiclassical approach has been developed for both inclusive and
diffractive structure functions at small $x$.
In this kinematic regime the momentum transfer to the proton,
needed to produce a diffractive state with invariant mass $M=O(Q)$,
is very small. Hence, it should be possible to describe deep inelastic
scattering as quark-antiquark pair production in a classical colour
field representing the proton. We have carried out such a
calculation based on a high
energy expansion for quark and antiquark wave functions in the presence
of a colour background field, which is assumed to be sufficiently smooth and
localized within some typical hadronic size. In the high energy expansion
corrections two orders beyond the leading eikonal approximation had to be
considered to obtain the complete leading order result for the cross section.
No expansion in the strong coupling constant $\alpha_s$ has been used.
Instead of gluons, triangular averages of the colour field are the
basic entities of the calculation.

The final result for the inclusive structure functions has been obtained
neglecting all terms suppressed by $x,\,\Lambda/Q$ or $\Lambda/M$. In
this limit $F_2$ and $F_L$ can be expressed in terms of two constants,
the actual value of which depends on the details of the proton field.
The inclusive structure functions are obtained in the limit $x\to 0$,
for which no unitarity problem exists. $F_2$ grows
logarithmically with $Q^2$, whereas the longitudinal structure function $F_L$
is independent of $Q^2$. This can be traced back to the
suppression of asymmetric configurations, with one relatively soft and one
hard quark. It is also interesting to observe, that the enhancement
of $F_2$ at large $Q^2$ is due to a term linear
in the field strength, i.e. this effect would survive an expansion in
$\alpha_s$.

The diffractive structure functions have been calculated from the
colour-singlet contribution to the above quark-antiquark pair production
cross section. Since the interaction with the proton is generally very soft,
it is natural to expect the appearance of a large rapidity gap event whenever
a colour neutral pair has been created. The colour-singlet projection in the
final state leads to the vanishing of the longitudinal structure function,
$F_L^D=0$, an effect connected with the already mentioned suppression of
asymmetric configurations.
Similar to the inclusive structure function, $F_2^D$ can also be expressed
in terms of two field dependent constants, given by integrals over certain
non-abelian phase factors testing the proton field. $F_2^D$ has no
$Q^2$-dependence and is given by some simple function of $\beta$ multiplied
with $\xi^{-1}$. The probably most important result of this paper
is the leading twist
behaviour of the diffractive structure function $F_2^D$. The exact
ratio of diffractive and inclusive cross-sections depends on constants,
sensitive to the details of the field, which have not been obtained
explicitly. At high $Q^2$ diffraction is found to be suppressed by a factor
of log$\,Q^2$ with respect to the total cross section.

In the last section we have argued that the difference between the $x$-slope
of $F_2$ and the $\xi$-slope of $F_2^D$ by one unit
is a generic feature of a large class of models.
This relation between $F_2$ and $F_2^D$ directly tests
the hypothesis of a common mechanism for diffractive and
ordinary deep inelastic scattering, with
the colour state of the proton remnant being responsible for the distinction
between the two event classes. The present calculation shows the possibility
of soft interactions adjusting the colour of the produced quark pair
as well as the
proton remnant to be in a singlet, resulting in leading twist
diffraction. In this sense the semiclassical approach, although quite
different from the partonic models \cite{heb,edin}, supports the
idea of soft interactions being responsible for diffraction in
deep inelastic scattering at small $x$.

Finally, it has to be mentioned that several important problems
require further study.  First, the very applicability of the
semiclassical approach to deep inelastic scattering
has not been established rigorously. Second, even if this could be done, it
would be necessary to integrate over the different field
configurations forming the proton. Together with a group
theoretical analysis, this could provide more information about the
field dependent functions introduced above. Another important issue,
not considered here, concerns hard gluon radiation in the process of
pair creation. We expect that this effect will significantly modify
the inclusive structure function at small $x$ as well as the
$\beta$-dependence of the diffractive structure function.\\[.1cm]

We would like to thank J. Bartels, M. Beneke, S. J. Brodsky, B. L. Ioffe,
E. Levin, M. L\"uscher and O. Nachtmann for valuable discussions
and comments. A.H. has been supported by the
Feodor Lynen Program of the Alexander von Humboldt Foundation.

\vspace{2cm}
\noindent
\\
{\bf\large Figure captions}\\
\\
{\bf Fig.1} Configuration space picture of quark-antiquark pair production
by a virtual photon in an external colour field.\\
\newpage
\thispagestyle{empty}
\vspace*{26cm}
\hspace*{-3cm}
\includegraphics{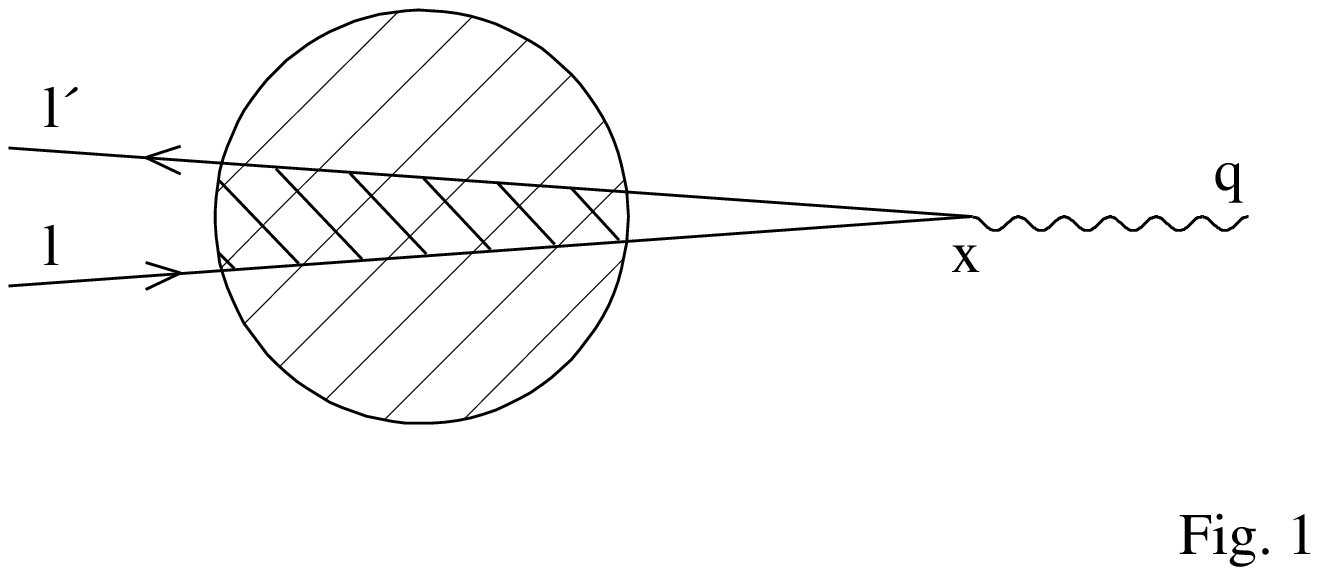}
\end{document}